# THz Metal-Mesh Bandpass Filters with Diamond-shaped Apertures for Sensitive THz Receivers

Changyun Yoo, Jeffrey L. Hesler, *Fellow, IEEE,* and Boris S. Karasik, *Senior Member, IEEE*

*Abstract*— We present high-performance terahertz (THz) metal-mesh bandpass filters developed to mitigate direct detection effects in sensitive THz receivers based on superconducting hot-electron bolometer (HEB) mixers. These metal-mesh filters are free-standing 5-µm thick sheets of copper perforated with a periodic array of diamond-shaped apertures. The simple aperture design minimizes the effect of fabrication (rounding) errors on filter performance, allowing precise engineering of the center frequency (1.5 - 5.5 THz) while maintaining high (>90 %) peak power transmission at normal incidence and a relatively narrow bandwidth of 5 – 15 %. Based on finite-element method (FEM) simulation results, we provide simple design equations that can be used for rapid design with high accuracy. The measured transmission profiles of 1.9-THz, 2.5-THz, and 4.7-THz filters show excellent agreement with the simulation results. These filters are compatible with cryogenic operation and can substantially reduce the direct detection effects in HEB mixers.

## I. INTRODUCTION

Terahertz (THz) bandpass filters are crucial components in sensitive THz receivers based on quasi-optically coupled superconducting hot-electron bolometer (HEB) mixers, where selective transmission of THz frequencies is necessary to calibrate receiver sensitivity accurately. Without these filters, the broadband thermal radiation from typical calibration loads at 295 K and 77 K can be significant enough to directly influence the electron temperature of the HEB mixer without the heterodyne mixing process taking place, thereby shifting the optimal bias point of the mixer and causing unwanted changes in the mixer gain and output noise [1-4]. This phenomenon, known as the "direct detection" effect, is particularly pronounced in quasi-optically coupled, small-volume HEB mixers with wide radiofrequency (RF) bandwidths, typically determined by the bandwidth of the THz antenna integrated with the mixer. For example, in a small-volume (2 µm × 0.2 µm × 4 nm) NbN-based HEB mixer integrated with a spiral THz antenna that has an RF bandwidth of ~ 2 THz, the difference in the power absorbed in the mixer device between the 295 K and 77 K loads is estimated to be ~ 3 nW (assuming a typical optical coupling loss of ~ 3dB), constituting a non-negligible (1.5 – 3 %) portion of the typical LO power requirement of 100 – 200 nW. For a voltage-biased HEB mixer, this direct detection effect is characterized by significant (~ 1 µA) shifts in the bias current when the hot and cold loads are switched during the calibration process, which can introduce 10 – 35 % errors in noise temperature measurements [1-3]. While several techniques, such as adjusting LO power in response to bias current shifts [5] or performing *ex-situ* analytical corrections [4], have been proposed to mitigate the direct detection effects, the former method can be inherently inaccurate due to the dependence of the mixer gain on the LO power. The latter method can be complicated and challenging to implement without accurately knowing the necessary input device parameters. As a result, using a narrow bandpass filter to limit the RF bandwidth of the receiver remains the most practical and straightforward solution to reduce the direct detection effects [1].

The most commonly used THz bandpass filters for this purpose are metal-mesh filters made from thin Cu or Au sheets with a periodic array of apertures that are either free-standing [6] or deposited on dielectric substrates [7-9]. While metal-mesh filters have been extensively studied in the context of frequency selective surfaces (FSS) in the microwave regime [10], designing and fabricating high-performance filters for THz frequencies (> 1 THz) has remained challenging. This is primarily due to losses in dielectric substrates that can be substantial at THz frequencies and the difficulties in fabricating the required small features with high precision. For example, THz metal-mesh filters on dielectric substrates typically exhibit peak power transmission ($T_p$) between 50 – 90 % with a relatively broad -3-dB bandwidth ($\Delta f_{-3dB}$) greater than 15 % of the center frequency $f_0$ at $f_0 > 1$ THz [7-9]. The best-performing filters are freestanding copper-mesh filters with cross-shaped apertures that offer $T_p > 95$ % and $\Delta f_{-3dB}/f_0 \sim 15$ % at $f_0 = 0.5 – 2.0$ THz [6]. However, cross-shaped apertures are highly sensitive to fabrication (rounding) errors, with typical rounding errors of 1 – 5 µm in the corners of the apertures resulting in 5 – 10 % shifts in $f_0$ at $f_0 \sim 2$ THz (see Section I of the Supplementary Material) and even more significant expected shifts at $f_0 > 2$ THz [11]. As a result, options for high-performance THz bandpass filters with $T_p > 90$ % and $\Delta f_{-3dB}/f_0 < 15$ % have remained limited above 2 THz, especially for applications in THz heterodyne receivers where precise alignment of $f_0$ with the targeted molecular emission lines is critical (see Table 1 for comparison of THz bandpass filters in the frequency range between 1.5 and 5 THz available from literature and commercial vendors).
-

Changyun Yoo and Boris S. Karasik are with the Jet Propulsion Laboratory, California Institute of Technology, Pasadena, CA 91109 USA (e-mail: chang.yoo@jpl.nasa.gov; boris.s.karasik@jpl.nasa.gov).
Jeffrey L. Hesler is with Virginia Diodes, Charlottesville, VA 22902 USA(e-mail: hesler@vadiodes.com).

TABLE I
COMPARISON OF THZ BANDPASS FILTER (1.5 – 5 THZ) FROM LITERATURE AND COMMERCIAL VENDORS

| Author / Vendor | $f_0$ (THz) | $\Delta f_{-3dB}/f_0$ (%) | $T_p$ (%) | Aperture Design | Substrate / Fabrication Process |
|---|---|---|---|---|---|
| Ma et al. [8] | 1.50 | 40 | 51 | Cross | HDPE / Photolithography & Lift-off |
|  | 1.75 | 51 | 54 |  |  |
|  | 2.91 | 14 | 52 |  |  |
| Porterfield et al. [6] | 1.52 | 19 | 97 | Cross | Freestanding / Photolithography & Electroplating |
|  | 2.08 | 15 | 100 |  |  |
| TYDEX [12][a] | 2.95 | 10 | 72 | Unknown | Freestanding / Unknown |
| QMC [13][a] | 1.90 | 22 | 80 | Unknown | Unknown / Unknown |
| Thorlabs [14][a, b] | 2.95 | 24 | 80 | Cross | Freestanding / Unknown |
| This work | 1.91 | 14 | 97 | Diamond-shaped | Freestanding / Electroforming |
|  | 2.50 | 15 | 93 |  |  |
|  | 4.70 | 12 | 98 |  |  |

a) Listed specifications correspond to standard filters for the selected frequency. Further customization for center frequency, bandwidth, and peak power transmission may be possible.
b) Currently discontinued. Specifications are based on the last available product datasheet.

This work introduces an improved version of free-standing copper-mesh filters featuring diamond-shaped apertures (Fig. 1(a)). Finite element method (FEM) simulation results using Ansys High-Frequency Simulation Software (HFSS) demonstrate that the diamond-shaped aperture is more robust against fabrication errors due to their simplicity in design (Fig. 1(b); also see Section II of the Supplementary Material for the details in HFSS modeling). This enables precise engineering of the bandpass profile up to ~ 5 THz while maintaining high peak power transmission and narrow -3-dB bandwidth, which has been challenging to achieve with traditional cross-shaped apertures. The diamond-shaped aperture can be designed using simple design equations and easily fabricated using commercially available electroforming techniques at low cost. Although these filters were initially developed for bandpass applications in THz receivers (and therefore evaluated only at normal incidence in this work), the simple design and high performance of these filters make them valuable for broader applications, including THz spectroscopy, imaging, and remote sensing, where selective transmission in the 2 – 5 THz range is essential.

## II. FILTER DESIGN

The bandpass performance for the diamond-shaped aperture is determined by two parameters – the diagonal length $J$ of the diamond aperture and the periodicity $G$ (Fig. 1(a)). Although the center frequency can be linearly scaled by independently adjusting $J$ (or $G$), fixing a constant $J/G$ ratio while scaling $J$ (or $G$) is critical for maintaining high peak power transmission and a narrow -3-dB bandwidth. As shown in Fig. 2, for a 5-µm thick copper-mesh filter, we found that a constant $J/G$ ratio between 0.6 and 0.7 guarantees a narrow -3-dB bandwidth ($\Delta f_{-3dB}/f_0 \approx 5 – 15$ %) and high peak power transmission ($T_p > 90$ %) for the entire range of center frequency ($f_0$ ~ 1.5 – 5.5 THz) tunable by scaling $J = 30 – 140$ µm.

In Figs. 2(a) and 2(b), each dot represents a simulation result corresponding to specific values of $J$ and $G$ (with fixed $J/G$ ratios). At the same time, the solid lines show smoothed curves, providing a clear visualization of the relative dependence of the filter performance on $J$ and $J/G$ ratios. The errors observed in each simulation result are less than 1 % and primarily stem from minor errors in $T_p$ limited by the given convergence tolerance and mesh size in the simulations. As shown in Figs. 2(a) and 2(b), the bandwidth of the filter is primarily determined by the relative size of the aperture (characterized by the $J/G$ ratio). In contrast, the peak power transmission scales with both the $J/G$ ratio and the absolute size of $J$. Consequently, achieving high peak power transmission is relatively more straightforward at the lower end of the 1.5 – 5 THz range with a broader -3dB bandwidth.

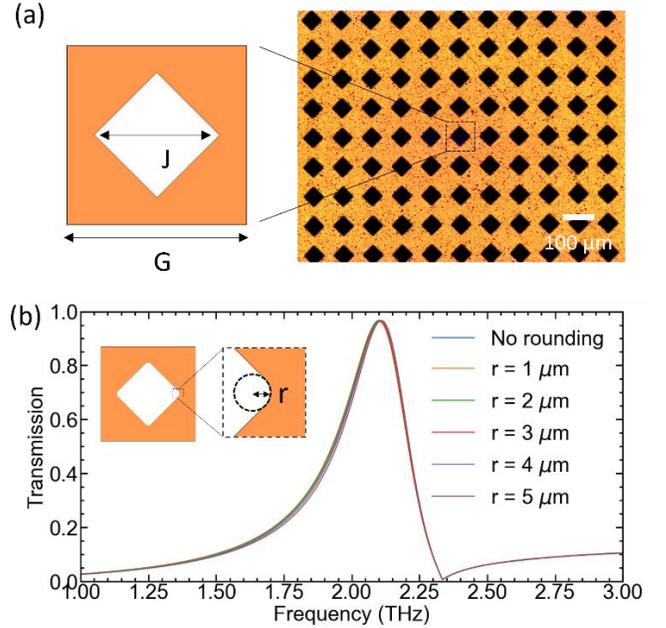

**Fig.1.** (a) Design for copper-mesh filters based on diamond-shaped apertures. (b) Simulated transmission profile of a 2-THz copper-mesh filter based on diamond-shaped apertures with varying degrees of rounding error $r$, demonstrating the robustness of the diamond-shaped aperture against fabrication errors.

The following simple design process determined our filters' design parameters $J$ and $G$. First, the required $J/G$ ratio is chosen based on the desired -3-dB bandwidth and peak power transmission according to the simulation results shown in Figs. 2(a) and 2(b). Next, the target center frequency determines $J$ and $G$ values from the design curve for the chosen $J/G$ ratio in Fig. 2(c). For example, for our 2.5-THz filter, we set $J/G = 0.7$ to achieve $T_p > 90$ % and $\Delta f_{-3dB}/f_0 \sim 15\%$ and determined $J = 76$ µm for $f_0 = 2.5$ THz using the design curve for $J/G = 0.7$ (brown curve in Fig. 2(c)) and $G = 108$ µm from the given $J/G$ ratio. Also, while the design curves in Fig. 2(c) can be directly used for the filter design, we found that the following design equation, acquired by retrofitting the design curves, can be used for rapid design with high accuracy in the frequency range between 1.5 – 5.5 THz:

$$f_0[THz] = \frac{300}{aJ[\mu m]} \quad (1)$$

where we have $a = -1.861\,J/G + 2.881$ for the diamond-shaped aperture. The design parameters for the 1.9-THz and 4-7 THz filters are provided in Table 2.

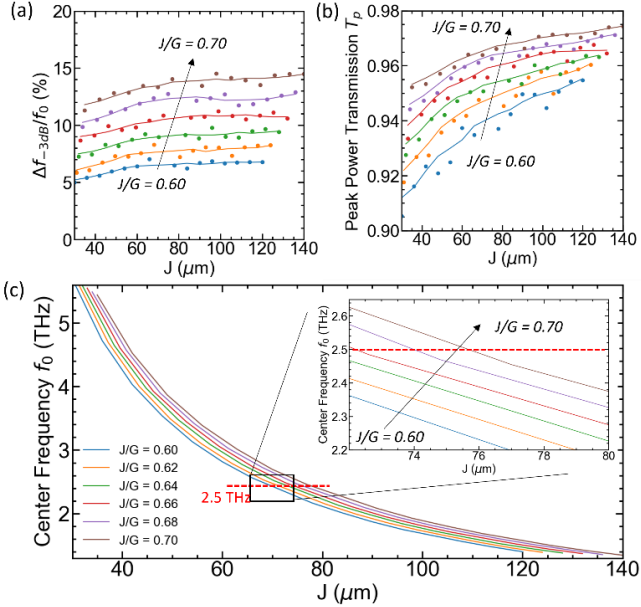

**Fig.2.** Simulated 5-µm thick copper filter performance based on diamond-shaped apertures. (a) -3-dB bandwidth as a function of $J$ for various $J/G$ ratios. (b) Peak power transmission as a function of $J$. (c) Center frequency as a function of $J$, with the inset showing a closer view of the design curves for filter design at 2.5 THz.

TABLE II
DESIGN AND FILTER PERFORMANCE OF COPPER-MESH BANDPASS FILTERS BASED ON DIAMOND-SHAPED APERTURES

| $J(\mu m) / G(\mu m)$ | | $f_0$ (THz) | | $\Delta f_{-3dB}/f_0$ (%) | | $T_p$ | |
|---|---|---|---|---|---|---|---|
| Design | Meas. | Design | Meas. | Design | Meas. | Design | Meas. |
| 99 / 142 | 97 / 142 | 1.90 | 1.91 | 14 | 14 | 0.97 | 0.97 |
| 76 / 108 | 76 / 108 | 2.51 | 2.50 | 14 | 15 | 0.97 | 0.93 |
| 40 / 58 | 40 / 58 | 4.70 | 4.67 | 12 | 12 | 0.95 | 0.98 |

### III. FABRICATION AND CHARACTERIZATION

The filters were fabricated from copper using a commercially available electroforming technique. In this technique, copper is deposited onto a patterned stencil via electrodeposition. Unlike electroplating, where the deposited metal adheres semi-permanently to the substrate, electroforming allows the deposited metal layer to be easily separated, resulting in a thin freestanding metal sheet that precisely follows the stencil pattern. This process enables the fabrication of thin (~3 µm) freestanding copper sheets with tight tolerances (<1 µm) maintained consistently over large areas (~70 cm²). In our case, the fabricated filters were 5 µm thick, each with a total area of 25 mm × 25 mm. While the individual patterns in the fabricated filters exhibited excellent uniformity, we observed a consistent overshoot of ~ 2 µm across all patterns, typical in the electroforming technique. This systematic deviation was accounted for by undershooting the design by the same amount. For example, for the 2.5-THz filter, the design pattern with $J = 74$ µm and $G = 104$ µm yielded the closest match to the intended design shown in Table 2. The measured dimensions of the fabricated filter were $J = 76 \pm 1$ µm and $G = 108 \pm 1$ µm, exhibiting excellent precision and uniformity over the entire area of the mesh.

The transmission profiles of the filters were characterized at normal incidence using a commercial Fourier-Transform Infrared (FTIR) Spectrometer. A mercury arc lamp was used as an unpolarized source, and a deuterated triglycine sulfate (DTGS) detector was used at room temperature. The frequency resolution was set to ~ 15 GHz. To limit light flux outside the target frequency range (1 – 6 THz), a Zitex G108 filter with a cutoff frequency of ~5.5 THz was placed in the sample compartment for reference and sample measurements. To prevent physical damage to the filters, we used a custom ring mount to handle the filters during the measurements (Fig. 3(a)).

As shown in Figs. 3(b-d), the measured transmission profiles (black solid line) of the 1.9-THz, 2.5-THz, and 4.7-THz filters showed excellent agreement with the simulated spectra (red dotted line). The designed and measured filter performance for these filters is also summarized in Table 2. A slight error (< 3 %) in the peak power transmission is attributed to the limitations in the dynamic range of the spectrometer, caused by the significant difference in the detector signal between the reference ("empty" spectrometer) and the filter measurements.

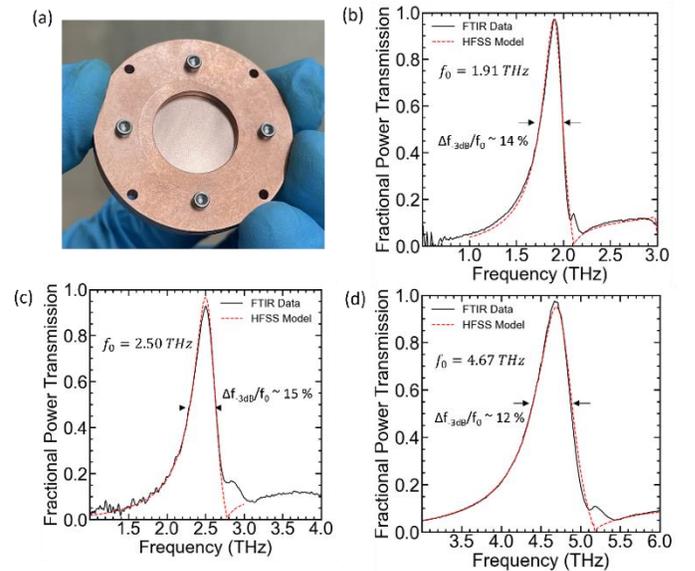

**Fig. 3.** (a) Custom ring mount for the filters. (b-d) Measured (black solid line) and simulated (red dotted line) transmission profiles for (b) 1.9-THz filter, (c) 2.5-THz filter, and (d) 4.7-THz filter.

To independently confirm the FTIR results, we conducted follow-up measurements for the 2.5-THz filter using the 2.522 THz line of a molecular gas far-infrared (FIR) laser with a pyroelectric detector. These measurements demonstrated $T_p \sim 95\%$, consistent with the high transmission observed in the FTIR measurements. Additionally, no dependence on polarization for the transmission profile was observed at normal incidence in FTIR and direct measurements.

Using these filters, the effective RF bandwidth of THz heterodyne receivers based on sensitive HEB mixers can be significantly reduced, thereby minimizing the direct detection effect during sensitivity calibration. For example, in THz receivers using NbN HEB mixers targeting 1.9 THz with a full RF bandwidth of ~ 2 THz, employing the 1.9 THz filter with a 3-dB bandwidth of ~ 270 GHz (as shown in Fig. 3(b)) reduces the effective bandwidth by a factor of 7.4. As a result, the difference in absorbed power between 295 K and 77 K calibration loads can be decreased to ~ 0.4 nW (compared to ~ 3 nW without the filter), assuming a typical optical coupling loss of ~ 3 dB, corresponding to only 0.2 – 0.4 % of the LO power requirement of 100 – 200 nW. These filters can be directly installed in front of the HEB mixers at the 4.2 K stage of the receiver cryostat, as their performance remains stable at cryogenic temperatures. The thermal contraction of copper is negligible compared to the typical dimensions of $J$ and $G$, leading to almost no impact on the center frequency and the -3-dB bandwidth.

Lastly, we note that, while we focused on diamond-shaped apertures in this work, other simple, wide aperture designs, such as circular apertures, can also serve as an alternative design for high-performance THz bandpass filters. Circular apertures, with their corner-free design, are inherently robust against fabrication errors and can be designed to achieve high peak power transmission ($T_p > 90\ \%$) and sufficiently narrow 3-dB bandwidth ($\Delta f_{-3dB}/f_0 \sim 5 – 15\ \%$) in the 2 – 5 THz frequency range (see Section III of the Supplementary Material for details).

## IV. Conclusion

In conclusion, we designed, fabricated, and characterized high-performance, free-standing THz copper-mesh bandpass filters with diamond-shaped apertures. These filters can demonstrate high peak power transmission (> 90 %) and narrow bandwidth (5 – 15 %) in a wide THz frequency range (1.5 – 5.5 GHz). The robustness of the diamond-shaped apertures against fabrication errors enables precise engineering of required bandpass performance with high accuracy. These filters are compatible with cryogenic operation and can substantially reduce the direct detection effects in THz receivers targeting 2 – 5 THz operation and in other applications where selective transmission of THz frequencies is essential.

## Acknowledgment

This research was carried out at the Jet Propulsion Laboratory, California Institute of Technology, under a contract with the National Aeronautics and Space Administration. Changyun Yoo's research was supported by an appointment to the NASA Postdoctoral Program, which Oak Ridge Associated Universities administers under contract with NASA.